\documentstyle[11pt,epsf,psfig,amsmath,axodraw]{article}

\def\Year{\expandafter\eatPrefix\the\year}
\newcount\hours \newcount\minutes
\def\monthname{\ifcase\month\or
January\or February\or March\or April\or May\or June\or July\or
August\or September\or October\or November\or December\fi}
\def\shortmonthname{\ifcase\month\orx
Jan\or Feb\or Mar\or Apr\or May\or Jun\or Jul\or
Aug\or Sep\or Oct\or Nov\or Dec\fi}

\def\TimeStamp{\hours\the\time\divide\hours by60%
\minutes -\the\time\divide\minutes by60\multiply\minutes by60%
\advance\minutes by\the\time%
${\rm \shortmonthname}\cdot   \if\day<10{}0\fi\the\day\cdot   \the\year
\qquad\the\hours:\if\minutes<10{}0\fi\the\minutes$}

\newskip\humongous \humongous=0pt plus 1000pt minus 100pt
\def\caja{\mathsurround=0pt}
\def\eqalign#1{\,\vcenter{\openup1\jot \caja
       \ialign{\strut \hfil$\displaystyle{##}$&$
        \displaystyle{{}##}$\hfil\crcr#1\crcr}}\,}
\newif\ifdtup

\newcounter{eqnumber}[section]
\renewcommand{\theeqnumber}{\thesection.\arabic{eqnumber}}
\def\equn{\refstepcounter{eqnumber}
\eqno({\rm \theeqnumber})
}

\def\Is#1#2{F
^{{#1}}_{4:#2}}
\def\Ione{\Is{\rm 1m}}

\def\Ihard{\Is{{\rm 2m}\,h}}

\newbox\charbox
\newbox\slabox
\def\s#1{{      
        \setbox\charbox=\hbox{$#1$}
        \setbox\slabox=\hbox{$/$}
        \dimen\charbox=\ht\slabox
        \advance\dimen\charbox by -\dp\slabox
        \advance\dimen\charbox by -\ht\charbox
        \advance\dimen\charbox by \dp\charbox
        \divide\dimen\charbox by 2
        \raise-\dimen\charbox\hbox to \wd\charbox{\hss/\hss}
        \llap{$#1$}
}}

\def\spa#1.#2{\left\langle#1\,#2\right\rangle}
\def\spb#1.#2{\left[#1\,#2\right]}
\def\lor#1.#2{\left(#1\,#2\right)}

\catcode`@=11  

\def\Slash#1{\hskip 0.05 cm \slash\hskip -0.22 cm #1}

\def\eps{\epsilon}

\def\e{\epsilon}

\def\half{{1\over 2}}

\def\la{\langle}
\def\ra{\rangle}

\def\lsl{\not{\hbox{\kern-2.3pt $\ell$}}}
\def\ksl{\not{\hbox{\kern-2.3pt $k$}}}

\def\spa#1.#2{\left\langle#1\,#2\right\rangle}
\def\spb#1.#2{\left[#1\,#2\right]}
\def\lor#1.#2{\left(#1\,#2\right)}

\def\sand#1.#2.#3{%
  \left\langle\smash{#1}{\vphantom1}\right|{#2}%
  \left|\smash{#3}{\vphantom1}\right\rangle}
\def\sandp#1.#2.#3{%
  \left\langle\smash{#1}{\vphantom1}^{-}\right|{#2}%
  \left|\smash{#3}{\vphantom1}^{+}\right\rangle}
\def\sandpp#1.#2.#3{%
  \left\langle\smash{#1}{\vphantom1}^{+}\right|{#2}%
  \left|\smash{#3}{\vphantom1}^{+}\right\rangle}
\def\sandmm#1.#2.#3{%
  \left\langle\smash{#1}{\vphantom1}^{-}\right|{#2}%
  \left|\smash{#3}{\vphantom1}^{-}\right\rangle}
\def\sandpm#1.#2.#3{%
  \left\langle\smash{#1}{\vphantom1}^{+}\right|{#2}%
  \left|\smash{#3}{\vphantom1}^{-}\right\rangle}
\def\sandmp#1.#2.#3{%
  \left\langle\smash{#1}{\vphantom1}^{-}\right|{#2}%
  \left|\smash{#3}{\vphantom1}^{+}\right\rangle}

\def\Atree{A^{\rm tree}}

\def\Lz{\mathop{\hbox{\rm L}}\nolimits_0}
\def\Kz{\mathop{\hbox{\rm K}}\nolimits_0}

\def\BR#1#2{\la#1^+|\Slash{K}|#2^+\ra}

\def\BRP#1#2#3{\la#1^+|\Slash{#2}|#3^+\ra}
\def\tree{{\rm tree}}

\def\NeqEight{{\cal N} = 8}
\def\NeqFour{{\cal N} = 4}
\def\NeqOne{{\cal N} = 1}
\def\NeqZero{{\cal N} = 0}

\begin{document}
\begin{titlepage}
\begin{flushright}
hep-th/0502028
\\
SWAT-05-425 \\
\end{flushright}

\vskip 2.cm

\begin{center}
\begin{Large}
{\bf One-Loop Gluon Scattering Amplitudes in Theories with
${\cal N} <4$ Supersymmetries}

\vskip 2.cm

\end{Large}

\vskip 2.cm
{\large
Steven J. Bidder,
N.\ E.\ J.\ Bjerrum-Bohr,
David C. Dunbar
and Warren B. Perkins}

\vskip 0.5cm

{\it
Department of Physics, \\
University of Wales Swansea,
\\ Swansea, SA2 8PP, UK }

\vskip .3cm

\begin{abstract}

Generalised unitarity techniques are used to calculate the coefficients
of box and triangle integral functions
of one-loop gluon scattering amplitudes in gauge theories
with ${\cal N} < 4$ supersymmetries.
We show that the box coefficients in $\NeqOne$ and $\NeqZero$ theories
inherit the same coplanar and collinear constraints as the corresponding $\NeqFour$ coefficients.
We use triple cuts to determine the coefficients of the triangle integral functions
and present, as an example, the full expression
for the one-loop amplitude
$A^{\NeqOne}(1^-,2^-,3^-,4^+,\ldots , n^+)$.

\end{abstract}

\end{center}

\vfill
\noindent\hrule width 3.6in\hfil\break
{\small \{pysb, n.e.j.bjerrum-bohr, d.c.dunbar,\ w.perkins\}@swan.ac.uk}
\hfil\break

\end{titlepage}

\section{Introduction}

The proposal of a ``weak-weak'' duality  between $\NeqFour$ super
Yang-Mills theory and a topological string theory~\cite{Witten:2003nn}
has led to
significant progress in the computation of amplitudes in gauge theories.

At tree level, amplitudes display a structure which is inherited from
the twistor string description.
This has inspired several
reformulations of tree level amplitudes.
Specifically, Cachazo, Svr\v{c}ek and Witten~\cite{Cachazo:2004kj}
proposed a formulation for calculating tree amplitudes using
``MHV-vertices'' rather than using conventional three and four point
Feynman vertices.  A MHV vertex is an off-shell continuation of the
Parke-Taylor formula~\cite{ParkeTaylor,Nair} for physical on-shell
tree amplitudes where two gluons have negative helicity and the
remaining helicities are all positive (these are
also known as ``Maximally Helicity Violating'' (MHV) amplitudes.)
This CSW formalism has proven very
useful in obtaining compact expressions for tree amplitudes and
has been extended to include external fermions and scalars~\cite{CSW:matter}
and even to theories with massive electroweak particles~\cite{Dixon:2004za}.
The MHV vertex approach extends to one-loop scattering
amplitudes as
demonstrated by the
re-computation of the MHV one-loop
amplitudes~\cite{Brandhuber:2004yw,QuigleyBedford,Bedford:2004nh}.

At one loop level, over many years, various techniques have been
developed to calculate loop-amplitudes more efficiently than
conventional Feynman diagram approaches.  A key ingredient is the
careful organisation of the amplitude in terms of the physical properties and
factorisation of the amplitudes.  (In fact, an important feature of
the CSW approach is that the MHV vertices are much closer to physical
amplitudes than Feynman vertices.)  Ideas such as the spinor helicity
formalism~\cite{SpinorHelicity} and colour-ordering~\cite{TreeColour},
which organise amplitudes according to the physical outgoing states
are very useful in determining tree amplitudes.  Beyond tree
level, the constraints demanded by unitarity have been used to compute
one-loop gluon scattering amplitudes in various supersymmetric
theories.  In $\NeqFour$ super Yang-Mills a one-loop amplitude is
completely specified by the coefficients of scalar box
functions~\cite{BDDKa,BDDKb}.  The one-loop MHV amplitudes have been
computed in both $\NeqFour$ super Yang-Mills~\cite{BDDKa} and in
$\NeqOne$ super Yang-Mills~\cite{BDDKb}.  The one-loop NMHV amplitudes
with three negative helicities and the rest positive (known as
next-to-MHV or NMHV amplitudes) have been calculated in $\NeqFour$
super Yang-Mills, first at six points~\cite{BDDKb}, then at seven
points~\cite{BeDeDiKo} and finally for all $n$~\cite{BDKn}. These
computations involve computing the two particle cuts~\cite{Cutkosky}
of an amplitude or more general cuts and factorisation
properties~\cite{GeneralisedCuts,BernChalmers}.

These methods have been complemented by techniques derived or
inspired by the twistor string approach.  MHV and NMHV tree amplitudes
have collinear and coplanar support in twistor space: these features
correspond to annihilation of the amplitude by particular differential
operators.  By acting with these differential operators on the cuts
of an amplitude one can
obtain~\cite{Cachazo:2004zb,Cachazo:2004by,Bena:2004xu,Cachazo:2004dr,Britto:2004nj}
algebraic equations which may be useful in computing the
box-coefficients in one-loop amplitudes. The utility of this approach
was demonstrated by the computation of one of the seven point
$\NeqFour$ one-loop amplitudes~\cite{Britto:2004nj}.
More recently, Britto, Cachazo and
Feng~\cite{Britto:2004nc} demonstrated, by continuing three-point
tree amplitudes to signature $(--++)$, how these box-coefficients could
be computed directly as a quadruple product of tree amplitudes.  (The
continuation of the signature can best be seen as a Lorentzian signature
with complex momenta. Although the unitarity properties are
obscure in normal field theory, the signature $(--++)$ is more natural
from a twistor space perspective~\cite{Witten:2003nn}.)

In this paper we examine generalised unitarity
techniques~\cite{GeneralisedCuts,BDKn} for calculating amplitudes in
theories with ${\cal N} < 4$ supersymmetries.  Firstly, we examine the
box-coefficients for a variety of helicity configurations in $\NeqOne$
and $\NeqZero$ theories: determining these from the quadruple
cuts~\cite{Britto:2004nc}.  These box coefficients satisfy
collinearity and coplanarity constraints which have a geometric
interpretation in twistor space.  Interestingly, the box-coefficients
obey these constraints independently of supersymmetry.  Specifically
the box-coefficients we compute are coplanar for NMHV amplitudes even
in the $\NeqZero$ case.

Box coefficients
are an important ingredient in these amplitudes but do not completely
specify the amplitude.
We demonstrate how triple cuts~\cite{GeneralisedCuts,BDKn} can be used to
determine the remaining triangle integrals and give the full
result for the previously unknown  amplitude,
$$
A(1^-, 2^-, 3^-, 4^+, 5^+, \ldots,   n^+ )\,,
\equn
$$
in the $\NeqOne$ theory.

\section{Generalised Unitarity and Relationships Between Box-Coefficients}

The idea that an amplitude might be reconstructed by its unitarity
constraints was originally investigated within the context of
$S$-matrix theories in the 1960's~\cite{Eden} with relatively limited
success.  However, these approaches assumed relatively little about
the actual theories considered. If one restricts these investigations to
theories which have a Quantum Field Theory description, {\it e.g.},
gauge theories, then these techniques have proven extremely useful. In
principle, a complete understanding of all cuts and factorisations in
all channels should be sufficient to completely reconstruct all loop
amplitudes.  Part of the complete understanding is that cuts must in principle be
evaluated with loop momentum in $4-2\epsilon$
dimensions.  However, for supersymmetric theories, amplitudes are
``cut-constructible''~\cite{BDDKa}, meaning that it is sufficient to calculate
the cuts using momenta restricted to four dimensions. This is an
enormous simplification, allowing one to exploit the relatively simple expressions
obtainable for on-shell tree amplitudes. While in special cases the two-particle cuts
are enough to compute an amplitude exactly, in other cases we must use higher-point and more
generalised cuts~\cite{BDDKb,GeneralisedCuts}. For example at two-loops one must also consider three
particle cuts and double-double cuts.

Within gauge-theories, amplitudes can be expanded in terms of
various integral functions,
$$
{\cal A} =  \sum \hat c_i I_4 +
\sum \hat d_i I_3 +\sum \hat e_i I_2  +\cdots\,,
\equn
$$
where, in general, theories with more supersymmetry have a
more restricted set of integral functions.  For $\NeqFour$ theories the series only
contains the scalar box functions, $I_4$, and hence is entirely
determined by the box-coefficients $\hat c_i$~\cite{BDDKa}.  For
$\NeqOne$ super Yang-Mills we have to consider  box functions
together with scalar triangle and bubble functions, $I_3$ and
$I_2$~\cite{BDDKb}.  For theories without supersymmetry the amplitude
may also contain rational pieces which have only been calculated in a
relatively small number of cases.

For $\NeqFour$ amplitudes analysis of the two particle cuts has
enabled a computation of the box-coefficients for arbitrary numbers of
particles in the  MHV~\cite{BDDKa} and NMHV cases~\cite{BDDKb,BeDeDiKo,BDKn},
either by evaluating the cuts or by acting on the cut with differential
operators~\cite{Cachazo:2004by,Cachazo:2004dr,Britto:2004nj}.

Recently, Britto, Cachazo and Feng demonstrated, by analytically
continuing tree amplitudes to a signature of $(--++)$, and using these
to calculate quadruple cuts, that box coefficients can be determined
algebraically from products of on-shell tree
amplitudes~\cite{Britto:2004nc}.  Specifically, considering a generic
amplitude containing the scalar box integral function,
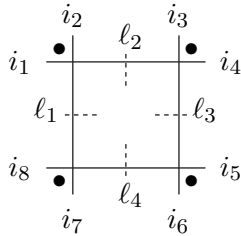
\vspace{-0,45cm}
\begin{center}
\begin{picture}(100,100)(0,0)
\DashLine(50,73)(50,61){2}
\DashLine(50,39)(50,27){2}
\DashLine(27,50)(39,50){2}
\DashLine(61,50)(73,50){2}
\Line(30,30)(30,70)
\Line(70,30)(70,70)
\Line(30,30)(70,30)
\Line(70,70)(30,70)

\Line(30,70)(20,70)
\Line(30,70)(30,80)

\Line(70,30)(70,20)
\Line(70,30)(80,30)

\Line(70,70)(70,80)
\Line(70,70)(80,70)

\Line(30,20)(30,30)
\Line(20,30)(30,30)

\Text(30,10)[c]{$i_7$}
\Text(10,30)[c]{$i_8$}
\Text(70,10)[c]{$i_6$}
\Text(90,30)[c]{$i_5$}
\Text(70,88)[c]{$i_3$}
\Text(90,70)[c]{$i_4$}
\Text(30,88)[c]{$i_2$}
\Text(10,70)[c]{$i_1$}

\Text(75,25)[c]{$\bullet$}
\Text(75,75)[c]{$\bullet$}
\Text(25,75)[c]{$\bullet$}
\Text(25,25)[c]{$\bullet$}

\Text(20,52)[c]{$\ell_1$}
\Text(52,20)[c]{$\ell_4$}
\Text(80,52)[c]{$\ell_3$}
\Text(52,80)[c]{$\ell_2$}
\end{picture}
\end{center}
\vspace{-0.5cm}
\noindent
its coefficient is given by
the product of four tree amplitudes where the cut legs
satisfy on-shell conditions,
$$\hspace{1cm}
\eqalign{
\hat c={ 1 \over 2 } \sum_{\cal S}
\biggl( \Atree(\ell_1,i_1,  & \ldots,i_2,\ell_2) \times
\Atree(\ell_2,i_3,\ldots,i_4,\ell_3)
\cr
& \hspace{2cm}\times \Atree(\ell_3,i_5,\ldots,i_6,\ell_4) \times
\Atree(\ell_4,i_7,\ldots,i_8,\ell_1) \biggr)\,,
\cr}
\equn\label{QuadEQ}
$$
where ${\cal S}$ indicates the set of helicity configurations and particle
types of
the legs $\ell_j$ giving a non-vanishing product of tree
amplitudes. The analytic continuation allows this to be evaluated even when one or more of the tree amplitudes in
eq.~(\ref{QuadEQ}) is a three point amplitude which would vanish in Minkowski signature.

In this section we restrict ourselves to a class of boxes where the
amplitude at each corner is either a MHV amplitude with two negative
helicity legs or a $\overline{\rm MHV}$ amplitude with two positive
helicity legs. This class of diagrams is quite large and includes all
helicity cases up to six-point amplitudes and the MHV loop amplitudes
themselves.  For convenience, we describe such amplitudes as
``MHV-deconstructible''.

We will consider three possible matter contributions to the box-coefficients;
the entire $\NeqFour$ multiplet; the $\NeqOne$ chiral
multiplet consisting of a fermion and a scalar; and the
contribution from a complex scalar circulating in the loop. We
often, perhaps perversely, describe these last as the $\NeqZero$ matter
contribution.
We can obtain the contribution of any matter
content by summing over linear combinations of these three matter
multiplets. Such decompositions arise very naturally in a string based approach~\cite{StringBased}.

For $\NeqOne$ super Yang-Mills with external gluons there are two
possible supermultiplets contributing to the loop amplitude ---
the vector and the chiral matter multiplets.  For simplicity we consider
colour-ordered one-loop amplitudes.
These can be decomposed into the contributions from single particle spins,
$$
\eqalign{ A_{n}^{\;\NeqOne\ {\rm vector}}\ \equiv\ A_{n}^{[1]}\
+A_{n}^{[1/2]} \,, \cr A_{n}^{\;\NeqOne\ {\rm chiral}}\ \equiv\
A_{n}^{[1/2]}\ +A_{n}^{[0]} \,, \cr}\equn
$$
where  $A_{n}^{[J]}$ is the one-loop amplitude with $n$ external
gluons and particles of spin-$J$ circulating in the loop.
(For spin-$0$ we mean a complex scalar.) For $\;\NeqFour$
super Yang-Mills theory there is a single multiplet
which is given by
$$
A_{n}^{\;\NeqFour}\ \equiv\
A_{n}^{[1]} + 4A_{n}^{[1/2]}+3 A_{n}^{[0]}\,.
\equn
$$
The contributions from these three multiplets
are not independent but satisfy
$$
A_{n}^{\;\NeqOne\ {\rm vector}}\ \equiv\ A_{n}^{\;\NeqFour} -
3A_{n}^{\;\NeqOne\ {\rm chiral}} \,.
\equn
$$%
Throughout we assume the use of a supersymmetry preserving
regulator~\cite{Siegel,StringBased,KST}.

We first
show that the box-coefficients for the three
matter contributions are {\it not} independent
for MHV-deconstructible boxes but that the $\NeqZero$ coefficient can be derived from the
$\NeqFour$ and $\NeqOne$ coefficients.
For MHV (and $\overline{\rm MHV}$ by conjugation) tree amplitudes the contributions
from the non-scalar particles can be related to that of the real
scalar via supersymmetric Ward identities~\cite{SWI,Nair}
and are simply,
$$
\Atree(( \ell_1)^{\mp}  ,i_1,   \ldots,i_2,(\ell_2)^{\pm} )
= (x)^{\pm 2h} \Atree(( \ell_1)^s ,i_1,   \ldots,i_2,(\ell_2)^{s} )\,,
\equn$$
where $h= 1/2$ for fermions and $h= 1$ for gluons
and $x =\spa{l_1}.{i_a}/\spa{l_2}.{i_a}$ with $i_a$ being the negative helicity gluon leg.
The contribution to the box-coefficient will then be
$$
(X)^{2h} \times {\rm real \; scalar \; contribution}\,,
\equn$$
where $X=x_1x_2x_3x_4$, and $x_j$ is the factor from the $j$-th corner.

When we consider the contribution from a supersymmetric multiplet to the
loop amplitude, we must sum over particle types.
For the
chiral multiplet the contribution, relative to the real scalar, has a factor
$$
\rho^{\;\NeqOne}  =  - X +2 -{1 \over X } =  -{ (X-1)^2 \over X}\,,
\equn$$
whilst for the $\NeqFour$ multiplet the factor is
$$
\rho^{\;\NeqFour} =  X^2 -4X +6 -4{ 1 \over X }
+{1 \over X^2}  = {(X-1)^4 \over X^2 } =
(\rho^{\;\NeqOne})^2\,.
\equn
$$
For $\NeqFour$ boxes we also have solutions where the two cut legs
attached to a corner have the same helicity. Such tree amplitudes are
only non-zero if the cut legs are gluons.  We refer to such configurations as
``singlet'' contributions.
It is the remaining ``non-singlet'' contributions which can be obtained from the scalar
by applying a factor of $\rho^{\; \NeqFour}$.
We thus have
$$
\hat c^{\;\NeqFour\; non-singlet} = \rho^{\;\NeqFour}\hat c^{\;real\;  scalar}
\;\;\;, \;
\hat c^{\;\NeqOne\; chiral} = \rho^{\;\NeqOne}\hat c^{\;real\;  scalar}\,,
\equn
$$
which given that $\rho^{\; \NeqFour}=(\rho^{\; \NeqOne})^2$ yields
$$
\hat c^{\; \NeqZero} = 2 {  ( \hat c^{\; \NeqOne\; chiral})^2
\over \hat c^{\; \NeqFour\; non-singlet}  }\,.
\equn
\label{magicEQ}
$$ This formula applies to any box which is MHV-deconstructible. It
can be used to determine the $\NeqZero$ (or scalar) coefficient from the two supersymmetric
coefficients {\it provided} we have identified the non-singlet
contribution in  the $\NeqFour$ case.

Such a formula will have several analogs in gravity amplitudes.
For graviton one-loop amplitudes
explicit formulations~\cite{GravityResults,BeBbDu} give,
$$
\hat c^{\; \NeqZero} = 2{  ( \hat c^{\; \NeqFour\; matter})^2
\over \hat c^{N=8\; non-singlet}  }\,,
\equn\label{magicGrav}
$$
where $\NeqEight$ denotes the
full $\NeqEight$ multiplet~\cite{ExtendedSugra},
$\NeqFour\; matter$ denotes the
$\NeqFour$ matter multiplet containing particles of spins $1$, $1/2$ and $0$
and ${\NeqZero}$ denotes the scalar contribution.

\noindent
Not all box-coefficients are MHV-deconstructible.
For example in the amplitude
$$
A(1^-,2^-,3^+,4^-,5^+,6^+,7^+)\equn
$$
the box
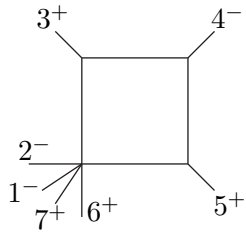
\vspace{-0.6cm}\begin{center}
\begin{picture}(100,100)(0,0)
\Line(30,30)(30,70)
\Line(70,30)(70,70)
\Line(30,30)(70,30)
\Line(70,70)(30,70)

\Line(30,30)(10,30)
\Line(30,30)(30,10)
\Line(30,30)(20,15)
\Line(30,30)(15,20)

\Line(70,30)(80,20)
\Line(70,70)(80,80)
\Line(30,70)(20,80)
\Text(13,86)[l]{$3^+$}
\Text(79,86)[l]{$4^-$}
\Text(80,15)[l]{$5^+$}
\Text(32,13)[l]{$6^+$}
\Text(6,36)[l]{$2^-$}
\Text(2,20)[l]{$1^-$}
\Text(12,10)[l]{$7^+$}
\end{picture}\hspace{3cm}
\end{center}%
\vspace{-0.4cm}
will have
a NMHV corner. The scalar tree amplitude at this corner is of the form
$$
{C_1 \over  K_{671}^2 }
+
{C_2 \over   K_{712}^2 }\,,
\equn
$$
where $K_{i\ldots j}\equiv (k_i+\ldots +k_j)$
and the amplitudes for other particles types~\cite{Tree,CSW:matter} are of the form
$$
x_1^h {C_1 \over   K_{671}^2 }
+
x_2^h {C_2 \over  K_{712}^2  }\,,
\equn
$$
which leads to box coefficients which are a sum of two terms
$$
\hat c =  \hat c_A + \hat c_B\,,
\equn
$$
each of which satisfy eq.~(\ref{magicEQ})
individually,
$$
 \hat c^{\; \NeqZero}_A = 2 {  (  \hat c^{\; \NeqOne \; chiral}_A )^2
\over \hat c^{\; \NeqFour\; non-singlet}_A  }
\;\;\; \hbox{\rm and}
\;\;\;
 \hat c^{\; \NeqZero}_B = 2 {  (  \hat c^{\; \NeqOne \; chiral}_B )^2
\over  \hat c^{\; \NeqFour\; non-singlet}_B  } \; .
\equn
$$
This formula has obvious generalisations to higher point box coefficients.

\section{Example Box Coefficients}
In this section we present some specific examples of ``MHV
deconstructible'' box-coefficients.  We use colour-ordered
amplitudes~\cite{TreeColour,Color} throughout and only present the
leading in colour expression.

There is a choice of representations for the box-integral functions.
There are scalar box-integral functions and $F$-functions which have
zero mass dimension and are related to the former by the removal of
the momentum prefactors~\cite{BDDKa},
$$
I_4 = { 1 \over K} F\,.
\equn
$$
We denote the coefficients of the scalar box functions as $\hat c_i$ and
those of the $F$-functions as $c_i$.
Both the $\hat c_i$ and $c_i$ satisfy the relations~(\ref{magicEQ}).

In all cases we present the $\NeqFour$, $\NeqOne$ and $\NeqZero$
results. For the $\NeqFour$ case the results are generally already
known~\cite{BDDKa,BDDKb,BeDeDiKo,BDKn} whilst the six point $\NeqOne$
box coefficients appear in~\cite{BBDPa}.

\subsection{MHV box-coefficients}

Consider the case of MHV-amplitudes where all box coefficients are known
and we may check the relationship~(\ref{magicEQ}).
 In general, the box functions are
``two-mass-easy'' boxes and single mass boxes.
The $\NeqFour$ non-singlet terms occur where there is a
single negative helicity leg in each massive corner.
The $\NeqFour$ amplitude
was calculated in ref.~\cite{BDDKa}
and the $\NeqOne$ in ref.~\cite{BDDKb} (the five point amplitude appeared earlier in~\cite{FiveGluon})
whilst the $\NeqZero$ coefficient was computed by Bedford et al \cite{Bedford:2004nh}.
Denoting the two negative helicities as $i$ and $j$ and considering the box with two
massless legs $m_1$ and $m_2$, the coefficients of the $F$-functions are
$$
\eqalign{
c^{\; \NeqFour} =& \Atree \times 1\,,
\cr
c^{\; \NeqOne}=& \Atree \times
{  b^{ij}_{m_1m_2} \over 2}\,,
\cr
c^{\; \NeqZero}=& \Atree \times
{  ( b^{ij}_{m_1m_2} ) ^2  \over 2}\,,
\cr}
\equn$$
where
$$
b^{ij}_{m_1m_2}
=
2{ \spa{i}.{m_1} \spa{i}.{m_2} \spa{j}.{m_1} \spa{j}.{m_2}
\over  \spa{i}.{j}^2 \spa{m_1}.{m_2}^2  }
\; ,
\equn
$$
and we use
spinor inner-products,
$\spa{j}.{l} \equiv \langle j^- | l^+ \rangle $,
$\spb{j}.{l} \equiv \langle j^+ | l^- \rangle $,
where $| i^{\pm}\ra $ is a massless Weyl spinor with momentum $k_i$ and
chirality $\pm$~\cite{SpinorHelicity,ManganoReview}.

\noindent
Clearly these amplitudes satisfy the relation~(\ref{magicEQ}).

\subsection{ Six point NMHV box-coefficients}

All boxes for the six point amplitudes are MHV-deconstructible and the
box coefficients are known for both $\NeqFour$ and
$\NeqOne$~\cite{BDDKb,BBDPa}, so we can apply (\ref{magicEQ}) to
generate the coefficients of the scalar boxes.  The amplitudes with
all-positive helicity legs and those with one-negative helicity leg
are non-zero in non-supersymmetric theories, however these amplitudes
are rational functions with no scalar box contributions.  Thus, the
two independent amplitudes with non-vanishing box-coefficients are the
MHV case (or $\overline{\rm MHV}$), which was covered in the previous
section, and the NMHV case with three negative helicities.

There are  three independent amplitudes with three negative helicity legs:
$A(1^-,2^-,3^-,4^+,5^+,6^+)$, $A(1^-,2^-,3^+,4^-,5^+,6^+)$
and $A(1^-,2^+,3^-,4^+,5^-,6^+)$.
Of these,
the first has vanishing box-coefficients for $\NeqOne$ and $\NeqZero$,
$$
\eqalign{
A^{\; \NeqZero , 1}(1^-,2^-,3^-,4^+,5^+,6^+)|_{{\rm box}} =
& \; 0\, .
\cr}\equn
$$
The $\NeqFour$ amplitude only has singlet contributions in this case.

\noindent
The second amplitude, $A(1^-,2^-,3^+,4^-,5^+,6^+)$,
does have a non-trivial box structure,
$$
\eqalign{
A(1^-,2^-,3^+,4^-,5^+,6^+)|_{{\rm box}} =&
\;
c_1^{} \Ihard{2;6}+c_2^{}\Ihard{2;2}+c_3^{}\Ihard{2;4}
+c_4^{}\Ione{5}+c_5^{}\Ione{6},
\cr}\equn
$$
which is depicted

\vskip 0.1 truecm
\begin{picture}(90,55)(0,0)

\Text(10,23)[c]{$c_1^{}$}
\Line(20,10)(60,10)
\Line(20,40)(60,40)
\Line(20,10)(20,40)
\Line(60,40)(60,10)

\Line(20,10)(20,0)
\Line(60,10)(60,0)

\Line(20,40)(15,50)
\Line(20,40)(25,50)

\Line(60,40)(55,50)
\Line(60,40)(65,50)

\Text(13,2)[l]{\small $5$}
\Text(7,50)[l]{\small $6$}
\Text(28,50)[l]{\small $1$}

\Text(48,50)[l]{\small $2$}
\Text(68,50)[l]{\small $3$}

\Text(63,2)[l]{\small $4$}

\end{picture}
\begin{picture}(90,55)(0,0)
\Text(5,23)[c]{$+\ c_2^{}$}
\Line(20,10)(60,10)
\Line(20,40)(60,40)
\Line(20,10)(20,40)
\Line(60,40)(60,10)

\Line(20,10)(20,0)
\Line(60,10)(60,0)

\Line(20,40)(15,50)
\Line(20,40)(25,50)

\Line(60,40)(55,50)
\Line(60,40)(65,50)

\Text(13,2)[l]{\small $1$}
\Text(7,50)[l]{\small $2$}
\Text(28,50)[l]{\small $3$}

\Text(48,50)[l]{\small $4$}
\Text(68,50)[l]{\small $5$}

\Text(63,2)[l]{\small $6$}

\end{picture}
\begin{picture}(90,55)(0,0)
\Text(5,23)[c]{$+\ c_3^{}$}
\Line(20,10)(60,10)
\Line(20,40)(60,40)
\Line(20,10)(20,40)
\Line(60,40)(60,10)

\Line(20,10)(20,0)
\Line(60,10)(60,0)

\Line(20,40)(15,50)
\Line(20,40)(25,50)

\Line(60,40)(55,50)
\Line(60,40)(65,50)

\Text(13,2)[l]{\small $3$}
\Text(7,50)[l]{\small $4$}
\Text(28,50)[l]{\small $5$}

\Text(48,50)[l]{\small $6$}
\Text(68,50)[l]{\small $1$}

\Text(63,2)[l]{\small $2$}

\end{picture}
\begin{picture}(90,55)(0,0)
\Text(3,23)[c]{$+\ c_4^{}$}
\Line(20,10)(60,10)
\Line(20,40)(60,40)
\Line(20,10)(20,40)
\Line(60,40)(60,10)

\Line(20,10)(20,0)
\Line(60,10)(60,0)

\Line(20,40)(20,50)

\Line(60,40)(50,45)
\Line(60,40)(70,45)
\Line(60,40)(60,45)

\Text(13,2)[l]{\small $3$}
\Text(13,50)[l]{\small $4$}

\Text(46,50)[l]{\small $5$}
\Text(58,50)[l]{\small $6$}
\Text(69,50)[l]{\small $1$}

\Text(63,2)[l]{\small $2$}

\end{picture}
\begin{picture}(90,55)(0,0)
\Text(5,23)[c]{$+\ c_5^{}$}
\Line(20,10)(60,10)
\Line(20,40)(60,40)
\Line(20,10)(20,40)
\Line(60,40)(60,10)

\Line(20,10)(20,0)
\Line(60,10)(60,0)

\Line(20,40)(20,50)

\Line(60,40)(50,45)
\Line(60,40)(70,45)
\Line(60,40)(60,45)

\Text(13,2)[l]{\small $4$}
\Text(13,50)[l]{\small $5$}

\Text(46,50)[l]{\small $6$}
\Text(58,50)[l]{\small $1$}
\Text(69,50)[l]{\small $2$}

\Text(62,2)[l]{\small $3$}

\end{picture}

\noindent
Of these coefficients, only three are truly independent, since under flipping, conjugation and relabeling,
$$
c_1 \leftrightarrow c_3 \; ,\;\;\;\;\;
c_4 \leftrightarrow c_5\,.
\equn$$
\noindent
Explicitly the remaining box-coefficients are,
$$
\begin{array}{llll}
c_1^{\; \NeqFour, \; non-singlet}
&=&\displaystyle i{ \BR31^4
\over
  \spb2.3\spb3.4\spa5.6\spa6.1 \BR25  \BR41 K^2
 }\,,
\cr
c_1^{\; \NeqOne \; chiral }
&=&\displaystyle i { \spa5.1 \BR31^2
\BR35
\over
 \spb2.3\spa5.6 \spa6.1
\BR25 \BR45^2},
&\hskip 0.7cm K=K_{234},
\cr
c_1^{\; \NeqZero}
&=&
\displaystyle 2 i
{\spa1.5^2 \spb3.4 \BR35^2  \BR41 K^2
\over
\spb2.3 \spa5.6\spa6.1 \BR25\BR45^4}\, ,
\cr
\end{array}
\equn
$$
$$
\begin{array}{llll}
c_2^{\; \NeqFour, \; non-singlet}
&=& \displaystyle i {  \BR34^4
\over
 \spb1.2\spb2.3\spa4.5\spa5.6
\BR14 \BR36 K^2  }\,,
\cr
c_2^{\; \NeqOne \; chiral}
&=& \displaystyle i
{ \spb3.1 \spa6.4\BR34^2
\over \spb1.2\spb2.3\spa4.5\spa5.6\BR16^2  },
&\hskip 1.0cm K=K_{123},
\cr
c_2^{\; \NeqZero}
&=&  \displaystyle 2 i
{\spb3.1^2\spa6.4^2 \BR14\BR36 K^2
\over
\spb1.2\spb2.3\spa4.5\spa5.6
\BR16^4}\, ,
\cr
\end{array}
\equn
$$
$$
\begin{array}{llll}
c_5^{\; \NeqFour, \; non-singlet}
&=&\displaystyle i { \BR64^4
\over
 \spb6.1\spb1.2\spa3.4\spa4.5
\BR63 \BR25 K^2}\,,
\cr
c_5^{\; \NeqOne \; chiral}
&=&\displaystyle i
{{ \BR64 } ^2
 \BR65
\over
 \spb6.1 \spb1.2
\spa3.5^2 \BR25 K^2}\,,
&\hskip 1.0cm K=K_{345},
\cr
c_5^{\; \NeqZero}
&=&\displaystyle 2 i
{\spa 3.4\spa4.5{\BR65}^2 {\BR 63} \over
\spa3.5^4\spb 6.1\spb1.2\BR 25 K^2}\, .
\cr
\end{array}
\equn
$$
The remaining amplitude, $A(1^-,2^+,3^-,4^+,5^-,6^+)$, contains all
six one-mass and all six ``two-mass-hard'' boxes,
$$\hspace{0.3cm}
\eqalign{
A(1^-,2^+,  3^-,4^+,  5^-, 6^+)_{\rm box}=
&
\sum_{i=1}^6 a_{i}^{}\Ione{i+3} +
\sum_{i=1}^6 b_{i}^{}\Ihard{2;i}
\cr} \equn
$$
These are not all independent and symmetry demands relationships
amongst the $a^{}_i$'s,
$$
\eqalign{
a^{}_3(123456)=a^{}_1(345612), \;\;\;
a^{}_5(123456)=a^{}_1(561234), \;\;\;
\cr
a^{}_4(123456)=a^{}_2(345612), \;\;\;
a^{}_6(123456)=a^{}_2(561234), \;\;\;
\cr
a^{}_2(123456) = \bar{a}^{}_1(234561), \;\;\;
a^{}_1(123456)=a^{}_1(321654),\; \;\;\;
\cr}\equn
$$
where $\bar a^{}_1$ denotes $a^{}_1$ with $\spa{i}.j \leftrightarrow \spb{i}.j$.
Thus there is a single independent $a^{}_i$. Similarly we can use symmetry to
generate all the $b^{}_i$'s from $b^{}_5$.  The
expressions for $a^{}_1$ and $b^{}_5$ are,
$$
\begin{array}{llll}
a^{\; \NeqFour, \; non-singlet }_1 & = &\displaystyle
i { \BR25^4
\over
 \spb1.2\spb2.3\spa4.5\spa5.6
\BR14 \BR36K^2
}\,,
\cr
a^{\; \NeqOne, \; chiral }_1 & = &
\displaystyle  i{ \BR{2}{5}^2 \BR15 \BR35 \over \spb1.3^2 \spa4.5 \spa5.6 \BR14\BR36 K^2  }\,,
&\hskip 0.5 truecm
K=K_{123},
\cr
a^{\; \NeqZero }_1 & = &
\displaystyle  2 i
{\spb1.2\spb2.3{\BR15}^2 {\BR35}^2 \over
\spb1.3^4\spa4.5\spa5.6{\BR14}{\BR36} K^2}\, ,
\cr
\end{array}
\equn
$$
$$
\begin{array}{llll}
b^{\; \NeqFour, \; non-singlet }_5 & =  &\displaystyle
i{ \BR25^4
\over
  \spb1.2\spb2.3\spa4.5\spa5.6
\BR14 \BR36 K^2}\,,
\cr
b^{\; \NeqOne, \; chiral }_5 &= &
\displaystyle i{ \BR{2}{5}^2  \BR35 \BR24
\over \spb1.2 \spa5.6 \BR36 \BR14 \BR34^2   }\,,
&\hskip 0.5 truecm
K=K_{123}.
\cr
b^{\; \NeqZero }_5 & = &\displaystyle
2 i {\spb2.3 \spa4.5\BR35^2\BR24^2 K^2
\over
\spb1.2 \spa5.6 \BR36\BR14 \BR34^4    }\,.
\cr
\end{array}
\equn
$$

\subsection{Two-Mass Hard Box}

As an $n$-point example, we can consider the coefficient of the
following box function,
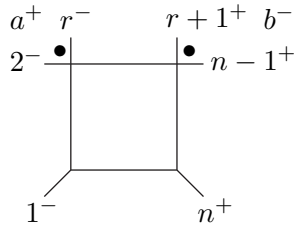
\vspace{-0.4cm}
\begin{center}
\begin{picture}(100,100)(0,0)

\Line(30,30)(30,70)
\Line(70,30)(70,70)
\Line(30,30)(70,30)
\Line(70,70)(30,70)
\Text(75,75)[c]{$\bullet$}
\Text(26,75)[c]{$\bullet$}

\Line(30,30)(20,20)
\Line(70,30)(80,20)

\Line(30,70)(20,70)
\Line(30,70)(30,80)

\Line(70,70)(70,80)
\Line(70,70)(80,70)

\Text(13,15)[l]{$1^-$}
\Text(78,15)[l]{$n^+$}
\Text(7,72)[l]{$2^-$}
\Text(26,88)[l]{$r^-$}

\Text(66,88)[l]{${r+1}^+$}
\Text(83,72)[l]{${n-1}^+$}

\Text(7,88)[l]{$a^+$}
\Text(103,88)[l]{$b^-$}

\end{picture}
\end{center}
\vspace{-0.6cm}

\noindent
which has two massless corners, a  corner with a single external
positive helicity leg and a corner with a single external negative helicity
leg. This box is thus MHV-deconstructible and can be computed using quadruple cuts
and the technique of Britto, Cachazo and Feng~\cite{Britto:2004nc}
whereby the massless legs are analytically continued to
signature $(--++)$ so that the massless corners do not vanish.

Solving for the box-coefficients we find
$$
\rho_{\; \NeqOne}
=
-{ \BR{1}{n}^2 \BR{a}{b}^2
\over
K^2\spb{a}.{1}\spa{n}.{b}
( K^2  \spb{a}.{1}\spa{n}.{b}-\BR{1}{n} \BR{a}{b} ) }\,,
\equn$$
where $K=K_{1\ldots r}$
and
the box coefficients
$$
\eqalign{
c^{\; \NeqFour \;  non-singlet}&  = i {
 s_{n1}
\BR{a}{b}^4
\over
\spb{1}.2  \ldots \spb{r-1}.r  \spa{r+1}.{r+2} \ldots
 \spa{n-1}.{n}
\BR{1}{r+1}\BR{r}{n} }\,,
\cr
c^{\; \NeqOne \; chiral} =& i
{   \spb{a}.{1}  \spa{b}.{n}
\Bigl(
K^2 \spb{a}.{1}   \spa{n}.{b}- \BR1{n} \BR{a}{b} \Bigr)   s_{n1} (K^2)
\BR{a}{b}^2
\over
\spb{1}.2 \ldots \spb{r-1}.r  \spa{r+1}.{r+2} \ldots
 \spa{n-1}.{n}
\BR1{n}^2
\BR{1}{r+1}\BR{r}{n} }\,,
\cr
c^{\; \NeqZero}=& 2i
{   \spb{a}.{1}^2  \spa{b}.{n}^2
\Bigl(
K^2 \spb{a}.{1}   \spa{n}.{b}- \BR1{n} \BR{a}{b} \Bigr)^2   s_{n1} (K^2)^2
\over
\spb{1}.2
\ldots
\spb{r-1}.r
\spa{r+1}.{r+2} \ldots
 \spa{n-1}.{n} \BR1{n}^4
\BR{1}{r+1}\BR{r}{n} }\,.
\cr}
\equn$$

\subsection{ The One-Mass Boxes}

For a one-mass box, adjacent massless legs must have opposite
helicity~\cite{Britto:2004nc} to yield a non-vanishing result upon
analytic continuation.
Using parity we
need only  consider the case where the massive corner is
mostly positive.
The case where exactly two of the massless legs
have positive helicity
is just the
MHV case considered previously.

The
remaining case where exactly two of the massless legs
have negative helicity
is a contribution to the NMHV amplitudes.
Specifically we have the one-mass scalar box:
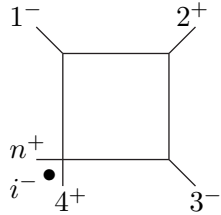
\vspace{-0.4cm}
\begin{center}
\begin{picture}(100,100)(0,0)
\Line(30,30)(30,70)
\Line(70,30)(70,70)
\Line(30,30)(70,30)
\Line(70,70)(30,70)
\Line(30,30)(20,30)
\Line(30,30)(30,20)
\Line(70,30)(80,20)
\Line(70,70)(80,80)
\Line(30,70)(20,80)
\Text(10,85)[l]{$1^-$}
\Text(73,85)[l]{$2^+$}
\Text(78,15)[l]{$3^-$}
\Text(27,15)[l]{$4^+$}
\Text(25,24)[c]{$\bullet$}
\Text(10,35)[l]{$n^+$}
\Text(10,20)[l]{$i^-$}
\end{picture}\hspace{3cm}
\end{center}
\vspace{-0.6cm}
Using the quadruple
cuts we can easily determine the coefficients in the three cases:
$$\eqalign{\displaystyle
c^{\; \NeqFour, \ non-singlet}
&=i{\BR2i^4\over  \spb1.2\spb2.3\spa4.5\ldots
\spa{i}.{i+1}\ldots\spa{n-1}.{n} \BR14\BR3n K^2}\,,
\cr
c^{\; \NeqOne \; chiral }
=i& {\BR2{i}^2\BR1{i}\BR3{i} \over
\spb1.3^2\spa4.5\ldots
\spa{i}.{i+1}\ldots\spa{n-1}.{n}\BR1{4}\BR3nK^2}\,,
\hskip 0.6cm K=K_{123},
\cr
c^{\; \NeqZero}=2&i{\spb1.2\spb2.3{\BR1i}^2 {\BR3i}^2 \over
\spb1.3^4\spa4.5\ldots\spa i.{i+1}\ldots\spa{n-1}.n
{\BR14}^2{\BR3n}^2 K^2}\,.
\cr}\equn
$$

\subsection{ The Two-Mass-Easy Boxes}
In the case of two mass easy boxes, there
are no solutions to the kinematic constraints if the
massless legs have opposite parity, so
$c^{\; \NeqZero}$, $c^{\; \NeqOne,\; chiral }$ and
$c^{\; \NeqFour,\; non-singlet}$ vanish for such
configurations. As an example of a non-vanishing
two mass easy box we consider the box below,
which has a single negative helicity leg at each corner.
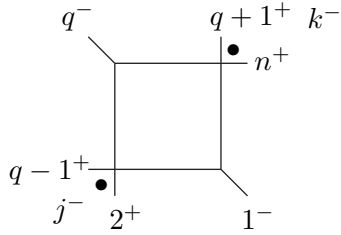
\vspace{-0.5cm}
\begin{center}
\begin{picture}(100,100)(0,0)
\Line(30,30)(30,70)
\Line(70,30)(70,70)
\Line(30,30)(70,30)
\Line(70,70)(30,70)
\Text(75,75)[]{$\bullet$}
\Text(25,24)[c]{$\bullet$}
\Line(30,30)(20,30)
\Line(30,30)(30,20)

\Line(30,70)(20,80)

\Line(70,30)(80,20)

\Line(70,70)(70,80)
\Line(70,70)(80,70)

\Text(78,12)[l]{$1^-$}

\Text(28,12)[l]{$2^+$}
\Text(7,15)[l]{$j^-$}
\Text(-10,30)[l]{$q-1^+$}

\Text(10,88)[l]{$q^-$}

\Text(66,88)[l]{${q+1}^+$}
\Text(83,72)[l]{${n}^+$}
\Text(103,88)[l]{$k^-$}

\SetWidth{1.0}

\end{picture}
\end{center}
\vspace{-0.6cm}
Setting, $K_2=k_2+k_3+..+k_j+..+k_{q-1}$ and $K_4=k_{q+1}+..+k_k+..+k_{n}$, we find,
$$
\begin{array}{llll}
c^{\; \NeqFour, \; non-singlet}
&=&\displaystyle
{i\over {\cal D}} \la j^-\vert K_2 K_4\vert k^- \ra^4,
\cr
c^{\; \NeqOne \; chiral }
&=&\displaystyle
-{i\over {\cal D}}
{
\BRP{q}{K_2}{j}
\BRP{1}{K_2}{j}
\BRP{1}{K_4}{k}
\BRP{q}{K_4}{k}
\la j^-\vert K_2 K_4\vert k^+\ra^2
\over
\spb{1}.{q}^2 },
&\hskip 1.0cm
\cr
c^{\; \NeqZero}
&=&\displaystyle
2 {i\over {\cal D}}
{
\BRP{q}{K_2}{j}^2
\BRP{1}{K_2}{j}^2
\BRP{1}{K_4}{k}^2
\BRP{q}{K_4}{k}^2
\over
\spb{1}.{q}^4 },
\cr
\end{array}
\equn
$$
where,
$$
\begin{array}{ll}
{\cal D}=
K_2^2K_4^2 &
\BRP{q}{K_2}{2}
\BRP{1}{K_2}{ q-1}
\BRP{1}{K_4}{q+1}
\BRP{q}{K_4}{n}
          \cr
&\hspace{2cm}
\times
\spa{2}.{3}\spa{3}.{4}..\spa{q-2}.{q-1}\spa{q+1}.{q+2}\spa{q+2}.{q+3}..\spa{n-1}.{n}.
\cr
\end{array}
\equn
$$

\section{Twistor Related Properties of Box Coefficients}
The results for the twistor structure of the box-coefficients are
relatively simple.
We find that the box-coefficients
within the MHV amplitudes have collinear support in twistor space
$$
F_{ijk} c^{\; \NeqFour\; MHV }=
F_{ijk} c^{\; \NeqOne\; MHV }=
F_{ijk} c^{\; \NeqZero\; MHV }=
0\,,
\equn$$
while box-coefficients within NMHV amplitudes
have coplanar support
$$
K_{ijkl} c^{\; \NeqFour\; NMHV }=
K_{ijkl} c^{\; \NeqOne\; NMHV }=
K_{ijkl} c^{\; \NeqZero\; NMHV }=
0\,,
\equn$$
in twistor space.
The coplanarity of the box-coefficients for the $\NeqFour$
amplitudes was shown in refs.~\cite{BeDeDiKo,Britto:2004tx}.
It was verified for the $\NeqOne$ box
coefficients in~\cite{BBDPa}.

In the generic NMHV case, where we have a three mass box, the legs will have support
upon three intersecting lines
in twistor space, with the legs at each massive corner being collinear.
The geometric picture of this is identical to that of $\NeqFour$~\cite{BDKn} and indeed
this pattern is also inherited by gravity amplitudes~\cite{BeBbDu}.
Since the three contributions in a supersymmetric decomposition obey the same
twistor space conditions,
it follows that
these conditions will apply
to gluon scattering
in many massless gauge
theories.

\section{Triangles from  Triple Cuts}
To obtain the coefficients of triangle integral functions we  consider triple cuts
~\cite{GeneralisedCuts}.
This
corresponds to inserting three $\delta(\ell_i^2)$ functions into the four
dimensional integrals. Specifically we consider,
$$
\eqalign{
\int d^4\ell_1d^4 & \ell_2d^4\ell_3
 \delta^4(\ell_1-\ell_2-K_1)\delta^4(\ell_2-\ell_3-K_2)
\delta(\ell_1^2)\delta(\ell_2^2)\delta(\ell_3^2)
\cr
& \times
\Atree(\ell_1,k_1,\ldots, k_r,\ell_2)
\Atree(-\ell_2,k_{r+1},\ldots, k_{r'},\ell_3)
\Atree(-\ell_3,k_{r'+1}, \ldots, k_n,-\ell_1)\,.
\cr}
\equn
$$ Both triangle functions and box functions contribute to this triple
cut. As a strategy, one can first determine the box-coefficients from
quadruple cuts and then subtract these from the triple cut to obtain
the triangle coefficients.  Unlike the quadruple cuts case, the three
$\delta(\ell_i^2)$ functions do not freeze the integral, so we must
carry out manipulations within the cut integral to recognise the
coefficient.

\noindent
As an example application of  triple cuts,
consider the amplitude
$$
A^{\; \NeqOne}(1^-,2^-,3^-,4^+,5^+, \ldots , n^+)\, .
\equn
$$
This amplitude is particularly amenable
in that it
contains no box integral functions.
This can be seen by examining the integrals in a two-particle cut~\cite{BBDD}
or, fairly obviously, by observing that there are no solutions to the
quadruple cuts.

\noindent
Consider the following triple cut:
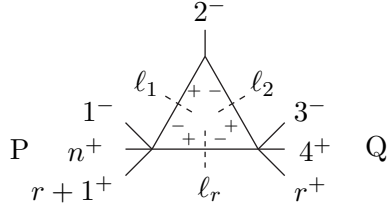
\vspace{-1.2cm}
\begin{center}
\begin{picture}(100,100)(0,0)


\Line(30,30)(70,30)
\Line(30,30)(50,65)
\Line(50,65)(70,30)

\Line(30,30)(20,40)
\Line(30,30)(20,20)
\Line(30,30)(20,30)

\Line(70,30)(80,40)
\Line(70,30)(80,20)
\Line(70,30)(80,30)

\Line(50,65)(50,75)

\Text(-20,30)[]{P}
\Text(115,30)[]{Q}
\Text(0,15)[]{${r+1}^+$}
\Text(5,30)[]{$n^+$}
\Text(10,45)[]{$1^-$}

\Text(90,15)[]{$r^+$}
\Text(92,30)[]{$4^+$}
\Text(90,45)[]{$3^-$}

\Text(52,83)[]{$2^-$}

\DashLine(35,50)(45,44){2}
\DashLine(65,50)(55,44){2}
\DashLine(50,23)(50,37){2}

\Text(28,55)[]{$\ell_1$}
\Text(72,55)[]{$\ell_2$}
\Text(52,15)[]{$\ell_r$}

\Text(46,52)[]{\tiny$ +$}
\Text(54,52)[]{\tiny$-$}

\Text(56,34)[]{\tiny$-$}
\Text(60,39)[]{\tiny$+$}

\Text(44,34)[]{\tiny$+$}
\Text(40,39)[]{\tiny$-$}

\end{picture}
\vspace{-0.6cm}
\end{center}
\noindent
with the momenta on the two massive legs being
$P\equiv k_{r+1}+\ldots +k_n+k_1$
and
$Q\equiv k_{3}+k_4+\ldots +k_r$.
Within the cut integral, where the cut legs are scalars,
the product of the three tree amplitudes
is
$$
{ \spa{1}.{\ell_1}^2 \spa{1}.{\ell_r}^2
\over
\spa{r+1}.{r+2}\ldots \spa{n}.1\spa1.{\ell_1} \spa{\ell_1}.{\ell_r}\spa{\ell_r}.{r+1}
 }
\times
{ \spa{3}.{\ell_2}^2 \spa{3}.{\ell_r}^2
\over
\spa{3}.{4}\ldots \spa{r-1}.{r}\spa{r}.{\ell_r}\spa{\ell_r}.{\ell_2}\spa{\ell_2}.{3}
}\times
{ \spa2.{\ell_1}\spa2.{\ell_2} \over \spa{\ell_1}.{\ell_2} }\,.
\equn$$
To obtain the contribution from the $\NeqOne$ multiplet we
must multiply this by  $\rho^{\; \NeqOne}$
within the integral.
Using
$$
{ 1\over \spa{\ell_1}.{\ell_r} }= { \spb{\ell_1}.{\ell_r} \over P^2 }
\;\;, \;
{ 1\over \spa{\ell_2}.{\ell_r} }= { \spb{\ell_2}.{\ell_r} \over Q^2 }\,,
\;\;
{\rm and}
\;\;
{ 1 \over \spa{r}.{\ell_r} } =
{ \spb{\ell_r}.2  \over \spa{r}.{\ell_r}\spb{\ell_r}.2  }
={ \spb{\ell_r}.2  \over
\BRP{2}{P}{r}
 }\; ,
\equn$$
this product can be rearranged to give
$$
{ F[ \ell_i ] \times \rho^{\; \NeqOne} \over
 \la 2^+ | P  | r^+ \ra
 \la 2^+ | P  | {r+1}^+ \ra \spa3.4\ldots \spa{r-1} .{r}
\spa{r+1}.{r+2}\ldots   \spa{n}.1 P^2Q^2  \spa{\ell_1}.{\ell_2}  }\,,
\equn$$
where much of the denominator can now be taken outside the cut integral and
$$
 F[ \ell_i ]
=\spa{1}.{\ell_1} \spa{1}.{\ell_r}^2
\spa{3}.{\ell_2} \spa{3}.{\ell_r}^2
\spa{2}.{\ell_1} \spa{2}.{\ell_2}
\spb{\ell_2}.{\ell_r} \spb{\ell_1}.{\ell_r}
\spb{2}.{\ell_r}^2\,.
\equn$$
When combining the different particles' contributions we have
$$
X = { \spa1.{\ell_1} \over  \spa1.{\ell_r}  }
\times { \spa2.{\ell_2} \over  \spa2.{\ell_1}  }
\times { \spa3.{\ell_r} \over  \spa3.{\ell_2}  }\,,
\;\;
{\rm so\  that}
\;\;
\rho^{\; \NeqOne} =
{  \Bigl( \spa1.{\ell_1}\spa2.{\ell_2}\spa3.{\ell_r}
- \spa1.{\ell_r}\spa2.{\ell_1} \spa3.{\ell_2}
\Bigr)^2
\over   \spa1.{\ell_1} \spa1.{\ell_r} \spa2.{\ell_2}\spa2.{\ell_1}
\spa3.{\ell_r} \spa3.{\ell_2}\,
} .
\equn$$
Thus the loop momentum dependent part of the integrand is
$$
{F[\ell_i] \rho^{\; \NeqOne} \over \spa{\ell_1}.{\ell_2} }={
\spa{1}.{\ell_r}
\spa{3}.{\ell_r}
\spb{\ell_2}.{\ell_r} \spb{\ell_1}.{\ell_r}
\spb{2}.{\ell_r}^2
\Bigl( \spa1.{\ell_1}\spa2.{\ell_2}\spa3.{\ell_r}
- \spa1.{\ell_r}\spa2.{\ell_1} \spa3.{\ell_2}
\Bigr)^2
\over \spa{\ell_1}.{\ell_2} }\,.
\equn$$
To evaluate this we use the identity
$$
\Bigl( \spa1.{\ell_1}\spa2.{\ell_2}\spa3.{\ell_r}
- \spa1.{\ell_r}\spa2.{\ell_1} \spa3.{\ell_2}
\Bigr)
=( \la 3^- |QP  | 1^+ \ra ) { \spa{\ell_1}.{\ell_2} \over \spb2.{\ell_r}}\,,
\equn
$$
which is valid due to
the momentum
constraints.
The part of the integrand which
still depends on the loop momentum can be rearranged
$$
\eqalign{
\spa{1}.{\ell_r}
\spa{3}.{\ell_r}
\spb{\ell_2}.{\ell_r} \spb{\ell_1}.{\ell_r}
\spa{\ell_1}.{\ell_2}
& =\spa{1}.{\ell_r}\spb{\ell_r}.{\ell_1}\spa{\ell_1}.{\ell_2}\spb{\ell_2}.{\ell_r}\spa{\ell_r}.{3}
\cr
&
= \la 1^- | \Slash{\ell_r}  \Slash{\ell_1}\Slash{\ell_2}\Slash{\ell_r} | 3^+ \ra
=
\la 1^- | \Slash{P}  \Slash{\ell_1}\Slash{\ell_2}\Slash{Q} | 3^+ \ra\,,
\cr}
\equn
$$
using $\Slash{\ell_r}=\Slash{\ell_1}+\Slash{P}$,
$\Slash{\ell_r}=\Slash{\ell_2}-\Slash{Q}$.
Finally we can reduce this to a linear function by using
$\Slash{\ell_1}=\Slash{\ell_2}+\Slash{k_2}$,
$$
\half
\la 1^- | \Slash{P}  \Bigl(
\Slash{k_2}\Slash{\ell_2}
-\Slash{\ell_1}\Slash{k_2}
\Bigr)
\Slash{Q} | 3^+ \ra\,,
\equn$$
where we chose to perform the algebra in such a way as to reflect the symmetry of
the diagram: this facilitates the identification of the triangle coefficients.
To solve this triangle we first Feynman parameterise
and make a shift of momenta
$$
\ell_1^\mu
\longrightarrow
\ell_1^\mu{}'
-k_2^\mu a_3  -(k_2+Q)^\mu a_{r+1}
\;\;\;\;
\ell_2^\mu
\longrightarrow
\ell_1^\mu{}'
-k_2^\mu a_3  -(k_2+Q)^\mu a_{r+1}
-k_2^\mu\,.
\equn
$$
leading to
$$
\half
\la 1^- | \Slash{P}  \Bigl(
\Slash{k_2}\Slash{Q}
-\Slash{Q}\Slash{k_2}
\Bigr)
\Slash{Q} | 3^+ \ra \times a_{r+1}\,.
\equn
$$
Finally, the Feynman parameter integral $I[a_{r+1}]$  can be expressed
in terms
of the  $\Lz$ functions
$$
I[a_{r+1}]
= {
\Lz [  P^2/ Q^2 ]
\over Q^2}\,.
\equn
$$
where we use the integral functions
$$
\eqalign{
\Lz [ r ] \ =&\ { {\rm ln} (r) \over 1-r } + {\cal O}(\e)\;  \,
\;\; {\rm and }\;\;\;
\Kz ( s) \ =\
\Bigl( {\rm -ln} (-s ) + 2 + {1\over\eps} \Bigr)+{\cal O}(\e)
\cr}.
\equn
$$
From the triple cut we can now identify the coefficient of the $\Lz$
triangle function:
$$
{ ( \la 3^- | QP  | 1^+ \ra )^2
\la 3^- | ( Q(2P-P2)P ) | 1+ \ra
\over
 \la 2^+ | P  | r^+ \ra
 \la 2^+ | P  | {r+1}^+ \ra \spa3.4\ldots \spa{r-1} .{r}
\spa{r+1}.{r+2}\ldots   \spa{n}.1 P^2Q^2 }\,.
\equn$$

Similarly,
we can determine all the triangle functions present in the amplitude
using triplet cuts,
obtaining the expression for the full amplitude
$$
\eqalign{
A^{\; \NeqOne}(1^-,2^-,3^-,4^+,5^+,\cdots, n^+)
= &
{\Atree \over 2} \left( \Kz( s_{n1} ) +\Kz( s_{34} )
\right)
-{i \over 2}
\sum_{r=4}^{n-1} \hat d_{n,r}
{    \Lz [ t^{[r-2]}_{3} / t^{[r-1]}_{2} ] \over t^{[r-1]}_{2}  }
\cr
&
-{i \over 2}
\sum_{r=4}^{n-2} \hat g_{n,r}
{    \Lz [ t^{[r-1]}_{2} / t^{[r]}_{2} ] \over t^{[r]}_{2}  }
-{i \over 2}
\sum_{r=4}^{n-2} \hat h_{n,r}
{    \Lz [ t^{[r-2]}_{3} / t^{[r-1]}_{3} ] \over t^{[r-1]}_{3}  }\,,
\cr}
\equn
$$
which can be depicted in the following way,
\vspace{-0.6cm}

\begin{picture}(100,100)(-210,0)
\Line(30,50)(65,70)
\Line(30,50)(65,30)
\Line(65,70)(65,30)

\Line(30,50)(20,60)
\Line(30,50)(20,40)
\Line(65,70)(75,70)
\Line(65,30)(75,30)
\Text(50,50)[]{$K_0$}
\Text(-120,50)[]{$A^{\; \NeqOne}(1^-,2^-,3^-,4^+,5^+,\ldots,n^+)$ \ \ = \ \ }
\Text(-10,50)[]{\ \ \ $\half {A^\tree}$}
\Text(85,71)[]{$1^-$}
\Text(85,32)[]{$n^+$}
\Text(23,50)[]{$\bullet$}
\Text(15,65)[]{$2^-$}
\Text(15,33)[]{$n-1^+$}

\end{picture}
\begin{picture}(100,100)(-230,0)
\Line(30,50)(65,70)
\Line(30,50)(65,30)
\Line(65,70)(65,30)

\Line(30,50)(20,60)
\Line(30,50)(20,40)
\Line(65,70)(75,70)
\Line(65,30)(75,30)
\Text(50,50)[]{$K_0$}
\Text(-20,50)[]{+\ \ $\half A^\tree$}
\Text(85,71)[]{$3^-$}
\Text(85,32)[]{$4^+$}
\Text(23,50)[]{$\bullet$}
\Text(15,65)[]{$2^-$}
\Text(15,33)[]{$5^+$}
\Text(115,50)[]{$+$}
\end{picture}
\vspace{-1.3cm}

\begin{picture}(100,100)(-40,0)


\Line(30,30)(70,30)
\Line(30,30)(50,65)
\Line(50,65)(70,30)

\Line(30,30)(20,40)
\Line(30,30)(20,20)
\Line(30,30)(20,30)

\Line(70,30)(80,40)
\Line(70,30)(80,20)
\Line(70,30)(80,30)

\Line(50,65)(50,75)

\Text(5,15)[]{${r+1}^+$}
\Text(10,30)[]{$n^+$}
\Text(15,45)[]{$1^-$}

\Text(90,15)[]{$r^+$}
\Text(92,30)[]{$4^+$}
\Text(90,45)[]{$3^-$}

\Text(52,83)[]{$2^-$}

\Text(-20,55)[]{$+\ \displaystyle\sum_{r=4}^{n-1} \hat d_{n,r}$}
\Text(50,43)[]{\small$[a_2]$}

\end{picture}
\begin{picture}(100,100)(-90,0)


\Line(30,30)(70,30)
\Line(30,30)(50,65)
\Line(50,65)(70,30)

\Line(30,30)(20,40)
\Line(30,30)(20,20)
\Line(30,30)(20,30)

\Line(70,30)(80,40)
\Line(70,30)(80,20)
\Line(70,30)(80,30)

\Line(50,65)(50,75)

\Text(15,15)[]{$2^-$}
\Text(10,30)[]{$3^-$}
\Text(15,45)[]{$r^+$}

\Text(90,15)[]{$1^-$}
\Text(92,30)[]{$n^+$}
\Text(90,45)[]{$r+2^+$}

\Text(52,83)[]{$r+1^+$}

\Text(-20,55)[]{$+\ \displaystyle\sum_{r=4}^{n-2} \hat g_{n,r}$}
\Text(50,43)[]{\small$[a_2]$}

\end{picture}
\begin{picture}(100,100)(-150,0)


\Line(30,30)(70,30)
\Line(30,30)(50,65)
\Line(50,65)(70,30)

\Line(30,30)(20,40)
\Line(30,30)(20,20)
\Line(30,30)(20,30)

\Line(70,30)(80,40)
\Line(70,30)(80,20)
\Line(70,30)(80,30)

\Line(50,65)(50,75)

\Text(15,15)[]{$3^-$}
\Text(10,30)[]{$4^+$}
\Text(15,45)[]{$r^+$}

\Text(90,15)[]{$2^-$}
\Text(92,30)[]{$1^-$}
\Text(90,45)[]{$r+2^+$}

\Text(52,83)[]{$r+1^+$}

\Text(-20,55)[]{$\displaystyle+\ \sum_{r=4}^{n-2} \hat h_{n,r}$}
\Text(50,43)[]{\small$[a_2]$}

\end{picture}\vspace{-0.4cm}\\
where,
$$
\eqalign{
\hat d_{n,r}= &
{ ( \la 3^- | {K}_{r-3}\overline{K}_{r-3}  | 1^+ \ra )^2
\la 3^- | {K}_{r-3}(k_2\overline{K}_{r-3}-\overline{K}_{r-3}k_2)\overline{K}_{r-3} ) | 1^+ \ra
\over
 \la 2^+ | \overline{K}_{r-3}  | r^+ \ra
 \la 2^+ | \overline{K}_{r-3}  | {r+1}^+ \ra \spa3.4\ldots \spa{r-1} .{r}
\spa{r+1}.{r+2}\ldots   \spa{n}.1 \overline{K}_{r-3}^2K_{r-3}^2 }\,,
\cr
\hat g_{n,r}= &
\sum_{i=1}^{r-3}
{\la 3^-|K_i \overline{K}_i|1^+\ra^2
\la 3^-|K_i\overline{K}_i(k_{r+1}\overline{K}_{r-3}-\overline{K}_{r-3} k_{r+1})|1^+\ra
\spa{i+3}.{i+4}
\over
\BRP{2}{K_i}{i+3}
\BRP{2}{K_i}{i+4}
\spa{3}.{4}\spa4.5\ldots\spa{n}.1 K_i^2 \overline{K}_i^2 } \,,
\cr
\hat h_{n,r}= &
(-1)^n \hat g_{n,n-r+2}\bigl\vert_{(123..n)\to(321n..4)}\,,
\cr}
\equn
$$
with
$
K_i=k_3+k_4+..+k_{i+3} $
and
$\overline{K}_i=k_2+k_3+..+k_{i+3}\,.$
We have checked that this expression satisfies the
correct collinear and soft limits thus confirming the normalisation.

\section{Conclusion}

Perturbative amplitudes in quantum field theories are complex objects
which contain a great deal of information, some of which is rather
well understood and some less so. The recently proposed relationships
between perturbative gauge theories and twistor strings provide a
fascinating insight into gauge theories and may be very useful in
perturbative calculations.  It also remains an open question as to whether
a string theory can be completely reconstructed from its states and its
on-shell tree amplitudes using unitarity and other techniques.

Although relations with twistor string theories have been
observed for $\NeqFour$ super-Yang-Mills, it is an open
question as to what degree theories with less or no supersymmetry are
related to a twistor string theory~\cite{Twistor}.
Until a direct connection is
uncovered it is reasonable to gather evidence by studying the
properties of amplitudes.
The box-coefficients are
a physically meaningful subset of an amplitude being the coefficients
of distinct functions of the class $\ln(s)\ln(s')$.
By computing some special examples, we have observed that even for non-supersymmetric
theories (but still massless) box-coefficients satisfy the same
collinearity and coplanarity constraints as in $\NeqFour$ theories.
These constraints can be seen as a consequence of the construction of
box coefficients using unitarity but may be a hint of the underlying
string structure.

For $\NeqFour$ theories the amplitudes are completely determined from
the box-coefficients. For theories with less supersymmetry the
amplitudes contain additional, and important, functional information.
As an example of using unitarity
constraints, we have presented the full structure of the simplest NMHV
configuration for $n$-gluons in $\NeqOne$ super-Yang-Mills.  This
amplitude is entirely expressed in terms of (specific) triangle
functions. The coefficients of these functions were determined by
carrying out triple cuts of the amplitude. These coefficients do not
have an obvious twistor property such as coplanarity.

Theories without supersymmetry are the most interesting
phenomenologically and, arguably, formally.  Unitarity techniques,
generalised sufficiently, may in principle determine such perturbative
amplitudes~\cite{DimShift,Bern:2005hs}
but practical computations are extremely
sparse at this point. It remains a challenge to develop techniques and perform
calculations for theories
without any supersymmetry.

\noindent{\bf Acknowledgements}\\
It is a pleasure to thank Zvi Bern and Nigel Glover
for useful discussions.  This work was supported by a PPARC rolling grant.
SB would like to thank PPARC for a research studentship.

\end{document}